\documentstyle[12pt]{article}
\begin{document}
\begin{center}
{
{\bf BLACK HOLES:  FROM GALACTIC NUCLEI TO ELEMENTARY 
PARTICLES}
{\footnote{ \noindent appeared in the Proceedings of 
the XXI$^{\rm th}$ SAB (Brazilian Astronomical Society) 
1995 Meeting, , edited by F. Jablonski, 
F. Elizalde, L. Sodr\'e Jr., V. Jatenco-Pereira, 
(IAG-USP 1996), p. 57.}}
} \\
\vskip 1mm
{\bf Jos\'e P. S. Lemos} \\
\vskip 3mm
{\scriptsize  Departamento de Astrof\'{\i}sica,
	      Observat\' orio Nacional-CNPq,} \\
{\scriptsize  Rua General Jos\'e Cristino 77,
	      20921 Rio de Janeiro, Brazil. \&} \\
{\scriptsize  Departamento de F\'{\i}sica,
	      Instituto Superior T\'ecnico,} \\
{\scriptsize  Av. Rovisco Pais 1, 1096 Lisboa, Portugal.}
\end{center}

\vskip 2mm
\begin{abstract}
\noindent
 
We present a broad review on black holes. We analyse some of the 
fundamental concepts in black hole theory, the observational and 
theoretical status of stellar and galactic black holes, and their 
appearance as quantum objects\end{abstract}
\noindent

\vskip 2mm
\noindent
{\bf 1. What is a black hole?}

\vskip 1mm

One of the basic ingredients of a physical theory is its set of 
fundamental constants. Thus, for instance, classical mechanics 
has no fundamental constants. The Newtonian theory of gravitation 
contains one constant alone, the universal constant of gravitation, 
$G$. The electromagnetism of Maxwell contains the velocity of light $c$, 
which in vacuum is a fundamental constant. Planck's constant $h$, which 
appeared directly from an experimental result (the black body 
spectrum), was immediately taken as the fundamental constant of 
quantum mechanics, developed among others by Bohr, Heisenberg and 
Schr\"odinger.  Thermodynamics yields the Boltzmann constant $k_B$ which 
arguably can be considered fundamental. 
These theories and their constants can then be combined 
to yield unified theories. If one tries to unite classical mechanics and
electromagnetism one obtains the theory of special relativity which 
has $c$ as a fundamental constant. The electric charge $e$ is a constant 
of nature. If we join $e$ and $c$ one obtains classical electrodynamics 
of Thomson and Lorentz, which is of course related to special relativity. 
If one further joins $h$ one obtains quantum electrodynamics due to Dirac 
and developed by others, afterwards. 
The unification of the electromagnetic and the weak forces 
by Weinberg and Salam as well as the inclusion of the strong force in 
the grand unified theories by Glashow and others also mix 
the fundamental constants of each separate theory. 
Finally, if one combines  $G$ and $c$ one obtains the theory of general 
relativity of Einstein. On further joining Planck's constant $h$ one 
should obtain quantum gravity, a theory which is still eluding the realm 
of physics, although there are some hints as to what it should be. 

Black holes (BHs) are objects which belong 
to the theory of general relativity, and
can be used to explain many powerful phenomena observed in the
celestial sphere. On the other hand, their (quantum) effects and
the singularities they hide yield an excellent framework to probe into
the nature of quantum gravity. 

A simple Newtonian argument can lead us to the concept of dark star, 
the Newtonian closest relative to the black hole (BH) of general relativity. 
The escape velocity $v_{e}$ of an object ejected from the surface of 
a body, such as a star, of mass $M$ and radius $R$, is given by 
$\frac12 v_{e}^2 =  \frac{GM}{R}$. 
The escape velocity on Earth is $11$ km/s, on a white dwarf it is around 
$6000$ Km/s. A dark star is defined as a star for which the escape 
velocity is greater or equal to the velocity of light, i.e., for 
which the following relation holds $\frac12 c^2 \leq \frac{2GM}{R}$. 
However, such a star is not a black hole for two reasons. 
First, the velocity of light 
is not a fundamental constant (i.e., it has no fundamental meaning) 
in Newtonian gravity, and thus other 
objects thrown with tachyonic velocities can escape and be detected at 
infinity. Second, the dark star is only dark for distant observers, near 
its surface the star is still bright since it emits light, although it 
cannot escape to infinity. 

The correct theory to explain the BH phenomenon is general
relativity. The BH is a ``state of the gravitational field'',
different from the state of the gravitational field of a star.  We can
understand BHs most easily through the collapse of a star 
(see e.g. \cite{MTW} \cite {lemos0}
\cite{lyndenbelllemos} \cite{lemoslyndenbell} \cite{lemos1}). As
the star collapses, its own radius shrinks. From  Newtonian gravity we
know that the force the star exerts on an object goes as $r^{-2}$.
Thus, a contraction by a factor two increases the force by a factor
four. In addition, if the star collapses to a point, the force becomes 
infinite at $r=0$. General Relativity yields a different result, the
gravitational force increases more rapidly than $r^{-2}$. The force is
then infinite when the radius of the star is $R=\frac{2GM}{c^2}$, the
Schwarzschild radius. The spherical surface formed at this radius is
called the event horizon. When a star of fixed mass $M$ attains this
radius, a BH is formed.  The difference between Newtonian
gravity and general relativity is of importance only when the star gets
closer to its Schwarzschild radius, where the gravitational field is
strong. Time near a strong gravitational field goes more slowly than
time far away, and space is highly curved. BHs are holes in
spacetime, caused by a strong space curvature and by drastic changes in
the flow of time.   

When the BH forms there are two regions connected to each other, 
the inside
and the outside of the event horizon. As the matter of the star
continues to collapse inside the event horizon it will form a
singularity where curvatures and densities of infinite strength are
formed.  Inside the event horizon light is trapped. Light  not only
does not escape to infinity, it cannot escape to the outside of the
BH.  However, to an outside observer the story is different.
As the BH is being formed, the luminosity of the original star decays
exponentially, $L=L_o e^{-\frac{t}{\tau}}$ where the characteristic
time is very short, $\tau=3\sqrt3\frac{GM}{c^3}= 2.6 \rm{x} 10^{-5}
\frac{M}{M_\odot}{\rm s}$.  In a few millionths of a second the star
turns totally black.  Another important feature is that the collapse of
the star results in a BH whose properties are characterized by
three parameters only:  mass, charge and angular momentum. One then
says that BHs have no hair (in fact, they have three hairs).
All the other properties, or ``hair'', of the matter of the star 
that formed the BH disappear.  No observation can reveal the nature of the
original star, whether it had a magnetic field, or possessed anti-matter,
or was made of fermions, or bosons, or it had any other hairs.

\vskip 3mm 

\noindent
{\bf 2. Stellar black holes}

\vskip 1mm

BHs with stellar masses can form through the collapse of the iron cores
of massive stars after they have reached the end of their thermonuclear
evolution. The outer layers of the star explode in a supernova leaving
at its center, depending on the core's mass, a neutron star or a BH. The
maximum mass for a neutron star is still a matter of debate, since it
depends strongly on the equation of state of the constitutive matter.
Rhoades and Ruffini \cite{ruffini} found a maximum mass of
$3.2M_\odot$, while Hartle \cite{hartle} can put an upper limit of
$5M_\odot$, (see also \cite{baym}). 
However, Bachall et al. \cite{bachall} argued that one can
construct a $100M_\odot$ 
star made of other types of matter at nuclear densities
which they called Q-stars (related in some sense to Witten's strange
stars \cite{witten1}). Thus the limit of the maximum mass $3.2$ or
$5M_\odot$ is still uncertain, although probably correct. The sizes 
of a BH and a neutron star do not differ much. For a $1M_\odot$ object, 
the neutron star has $10$ Km of radius, whereas the BH has $3$ Km. 

An isolated stellar BH cannot be seen. A BH can only be observed if it
belongs to a binary system \cite{michel}, and is  detected through
spectroscopic observations of the bright optical companion.  The main
problem is that the unseen body can also be a neutron star, and to
distinguish between both possibilities one has to follow 
a complicated list of steps.  A
binary system can evolve in the following way: first, two massive
stars, with masses of the order of $20M_\odot$ form a binary system.
Then, in a second stage, 
the more massive star evolves more rapidly and soon becomes a compact
body with a few $M_\odot$ after having exploded in a supernova.  
Finally, 
the other star also evolves to become a similar compact star. One
thus has a binary system of two compact stars, of which the most famous
example is the binary pulsar of Hulse and Taylor \cite{taylor}. During
the second stage, when the binary is composed of one compact and one 
giant star there
is the production of spectacular phenomena visible in the X-ray band. The
binary systems in this stage are called X-ray binaries.  

X-ray binaries have a very short orbital period which by Kepler's third
law implies the objects are very close. Since the Roche lobe of the
binary system (the surface of gravitational neutrality) can be 
filled in part
by the massive companion, the outer layers of the massive 
star are captured by
the compact star. The captured gas then forms an accretion disk. The
emission of X-rays can happen through several processes. If the compact
star is a neutron star then the gas, through a magnetic field
mechanism, hits the crust of the neutron star regularly with the
consequent emission of X-rays. The neutron star is then called an X-ray
pulsar, with luminosities of the order $10^4 L_\odot$. If, instead of
regular, the X-ray emission is sporadic, then the source is called a
burster. Bursters are usually produced through an explosion of the
surface of the neutron star, but can also appear by eruption of a very
hot region of the accretion disk \cite{shapteuk} \cite{luminet}. 
In this last case the matter from the luminous companion spirals 
towards the unseen compact object and emits X-rays with temperatures 
of  $10^8-10^9$K. These temperatures are generated through dissipation 
by viscous processes of the gravitational energy of the infalling 
matter accelerated to high velocities.

BHs do not have a hard solid surface, the explosion happens in 
the disk. Is there any way to distinguish between BH and 
neutron stars bursts? One could think that variability would give some 
clues.  If a source (disk, in the case) changes shape, the speed of 
change cannot exceed the speed of light. If one detects variability 
in a time $\Delta t$, the size of the source is at most 
$l \buildrel<\over\sim c\Delta t$. If the source changes in 
$\Delta t \sim 10^{-3}$s, then 
its linear size is $l\sim 300$ Km. For a stellar BH, its inner edge 
(defined as the last stable orbit, see e.g. \cite{andlemos}), 
is of the order 
of $30$ Km in radius say, so its circumference is around 
$200$ Km. A rotating hot bubble emitting X-rays would have a varability 
in the milisecond range, as observed. This could be a signature for 
a BH. However, Circinus X-1 also shows fluctuation of this order, 
and it was shown that it also has periodic bursts which characterizes a 
neutron star \cite{ref1}. The identification of a black hole through 
radiation processes is not yet well developed, although it is a field 
advancing quickly. The best criterion to identify a BH is to find 
its mass through dynamical studies of the X-ray binaries. The weighing 
of stars in binaries is a technique fully understood nowadays. 

Knowing the orbital period of the binary and the projected mean speed 
of the optical star, one can using Kepler's law to deduce the mass 
function defined by \cite{smith},
$f(M) \equiv \frac{(M_x \sin i)^3}{(M_x+M_c)^2}
=\frac{P(V_{c}\sin i)^3}{2\pi G}$, 
where $M_x$ and $M_c$ are the masses of the X-ray source and the companion 
respectively, $i$ is the orbital inclination angle, $P$ is the 
orbital period and $V_c \sin i$ is the projected velocity semiamplitude of 
the optical companion. If one puts $M_c=0$ and $i=90^{0}$ one gets the 
mimimum possible value for $M_x$ which in this case is equal to 
$f(M)$. What one would really like to obtain is an $f(M)$ greater 
than $5M_\odot$. However, following the theory, a value close to 
$3.2M_\odot$ yields already a good BH candidate.

There are three very strong candidates, two good candidates and a list
of possible candidates. We start here giving some properties of the
three very strong candidates. {\bf (1a) Cygnus X-1} -- It was discovered
in 1971 \cite{hjelming} and it is a persistent source with $L_x \sim 2
{\rm x} 10^{37}$erg/s. It is a high mass X-ray binary.  The
orbital period is 5.6 days and $V_c \sin i \simeq 76$Km/s which yields
a low $f(M)=0.35M_\odot$. Now, one has to derive a reliable lower limit
for the mass, which is a difficult task. The optical companion is a
blue giant with mass $M_c \sim 30 M_\odot$. The inclination angle is
supposed to be $i \sim 30^0$ (there are no eclipses). This gives a mass
of $M_x\sim 16 M_\odot$ \cite{liangnolan}. The most conservative
assumptions lead to $M> 3M_\odot$.  {\bf (1b) LMC X-3} -- It was
discovered in 1983 \cite{cowley1}. It is a persistent source and a
high mass X-ray binary with $L_x  \sim 3 {\rm x} 10^{38}$erg/s. 
$P=1.7$ days and $V_c
\sin i \simeq 235$Km/s. This gives $f(M)=2.3M_\odot$. The mass of the
companion is estimated to be $M_c\sim 6M_\odot$ which then yields
$M_x\sim 6M_\odot$. Since the distance to the Magellanic cloud is known
one can use Paczy\'nski method \cite{paczyn1} to infer 
$M_x \buildrel>\over\sim  4M_\odot$.  {\bf (1c) 0620-00} -- It was
discovered in 1986 \cite{mclintock1}. Contray to the other two, 
it is a transient source. It is a
low mass X-ray binary with $L_x  \sim 1 {\rm x}
10^{38}$erg/s.  $P=0.32$ days and $V_c\sin i \simeq 467$ Km/s yielding
$f(M) = 3.18M_\odot$. Based solely on the value of the mass function,
the minimum mass is already equal to the BH threshold mass. It is
considered the strongest of the very strong candidates. $M_c\sim
0.7M_\odot$ which is low, and yields the following lower limit $M_x
\sim 4M_\odot$. 

There are two other candidates which have been weighed, although the
uncertainties are greater than the sources mentioned above.  {\bf (2a)
CAL87 } --  One has $L_x  \sim 1 {\rm x} 10^{36}$erg/s. It is an
interesting system since it undergoes eclipses. If $M_c >0.4 M_\odot$ it was
found that $M_x \buildrel>\over\sim 4M_\odot$ \cite{cowley2}.  
{\bf (2b)
LMC X-1} --  $L_x  \sim 2 {\rm x} 10^{38}$erg/s. The optical companion
is still not identified conclusively. However, there are hints that
$M_x \buildrel>\over\sim 3M_\odot$.  There are a number
of other candidates which have been selected because they show X-ray
behavior similar to Cygnus X-1, of which the prime example is GX239-4,
and others which show transient behavior similar to 0620-00, such as
GS2000+25, GS2023+33, GS1124-68, 4U1543-47, 4U1630-47, H1705-250
\cite{cowley3}.  Further dynamical studies are needed to obtain the masses
of these sources . The spectacular source SS433 which shows
emission of jets was recently discarded as a black hole, since its mass
was shown to be $M\sim 1.4M_\odot$ \cite{ref2}.

What are then the features that allow us to identify a BH candidate? 
The classical steps are: 1) The luminosity of the X-ray source has 
to be high  $L_x  > 10^{36}$erg/s and of rapid variability $<1$s.  
This implies that the binary system must contain an accreting  
compact object. 
2) The optical companion is identified and allows to measure 
the orbital period and the projected orbital velocity, to yield $f(M)$. 
3) The mass of the optical companion and the inclination 
of the orbit are inferred or limited, based on the distance, optical 
spectrum, $L_c$, and eclipes. Then using $f(M)$ one deduces $M_x$.
4) If the mass of the object is $M_x\buildrel>\over\sim 3.2M_\odot$ 
then it is considered a BH candidate. There are now three other criteria 
which can help in identifying a BH candidate: (i) the source has a 
spectrum with soft X-rays, $\sim 1$Kev. (ii) $Fe$ emission lines very 
near the compact object (will) allow one to measure the 
velocity of the rotating disk.  
This then implies a dynamical measure of $M_x$ \cite{fabianstella}.
(iii) Hard X-rays $\sim 100$Kev are a signature of BHs, since neutron stars 
have a hard surface whose radiated photons cool the accretion disk through 
Compton cooling \cite{sunayev}. These last too criteria are very recent 
\cite{paradijs1}, and it is expected the situation will improve with work 
on some other half-dozen sources. 

One drawback is that all these criteria are indirect. One really wants to 
come close to the collapsed object.  In the long run, one is after clear 
evidence for the existence of an event horizon \cite{mclintock2} 
\cite{stella}. This might be possible after the gravitational 
antennas are fully operating, to detect unambiguously the formation of 
a BH. If the BH is in a binary one expects subsequent X-ray 
emission. 
How many BHs there are in the Galaxy? The last 
estimates give 1000-3000 BHs, of the same order 
as the number of neutron stars \cite{cowley3}. 
\vskip 3mm

\noindent
{\bf 3. Black Holes in Galactic Nuclei}

\vskip 1mm

We have seen that in the complete gravitational collapse of a star a BH
can form with mass in the range $3-20M_\odot$. Yet, the theory of
gravitational collapse allows for the formation of BHs with much
greater masses, masses that can be in the range $10^3-10^9M_\odot$.
These BHs may appear in the core of clusters of stars or in the center
of galactic nuclei. If the mass of the original system is very large,
there is no uncertainty in the equation of state of the collapsing
matter when it crosses the event horizon.  Indeed, at
$R=\frac{2GM}{c^2}$, the density of the matter is $\rho =
\frac{3c^6}{32\pi G^3}\frac1M^2\simeq 1.3 {\rm x}10^{16}
(\frac{M_\odot}{M})^2$g/cm$^3$. For a $\sim 1M_\odot$ BH the density is
very high, above the nuclear density. However, for a $\sim 10^8M_\odot$
object one has $\rho \sim 1$gm/cm$^3$. In this case one has
$\frac{R^3}{{R_\odot}^3} \sim 10^8$. This roughly means that for a
cluster composed of $10^8$ suns, the cluster crosses its own
Schwarzschild radius when the suns, uniformly distributed over the
volume,  are touching each other. For a $\sim 10^{10}M_\odot$ object
one has $\rho\sim 10^{-4}$gm/cm$^3$. In this case,
$\frac{R^2}{{R_\odot}^2} \sim 10^{10}$, which roughly means that for a
cluster made of $10^{10}$ suns, it crosses its Schwarzschild radius
when the suns, distributed uniformly over a spherical layer of
thickness of one sun diameter, are touching each other. In all these
latter cases the physics when the matter crosses the horizon is well
known. 

Theory and numerical simulations favor the appearance of a binary system 
in the center of globular clusters. The compact binary scatters any 
incoming star. There is, in principle no formation of a BH in the 
core of the cluster. This is supported by observation. However, central 
BHs are not totally excluded \cite{shapteuk}. 

There is controversy about the existence of a central BH in our Galaxy
since it was first proposed in 1971 \cite{dlb1}. The Galactic center
has the following features: (1) a disk of gas with inner and outer
radii given by $5-30$ light years (ly); (2) a cavity interior to 
the disk with $2{\rm x}10^6$ stars; 
(3) a possible BH with $2{\rm x}
10^6{\rm Km} \sim 2{\rm x} 10^{-7} {\rm ly}$ accreting matter slowly.  The
evidence for a central compact source comes from the radio emission of
a region as small as the orbit of saturn around the Sun \cite{lo}. This
source is called Sgr A$^*$ and has $L\sim 10^{34}{\rm erg/s} 
\sim 10L_\odot$ ($\sim 10^4$ times the luminosity of a 
single radio pulsar). From the disk of gas, one can infer (if 
it is in a Keplerian orbit) a central mass of 
$5-8{\rm x}10^6M_\odot$. Subtracting 
the mass in the red giants one obtains  $3-6{\rm x}10^6M_\odot$. The 
evidence favors the existence of a central 
BH, although it is not absolutely convincing 
\cite{genzel1} \cite{genzel2}. Evidence against the existence of a central 
BH has appeared after observations 
from the Sigma/GRANAT telescope  led to the conclusion that, 
contrary to expectations,  
there is no X-ray source coincident with Sgr A$^*$ \cite{goldwurm}. 
However, there are now models which can explain the phenomenon in 
a natural way in which the matter is swallowed before it has 
time to radiate \cite{narayan}. 

There is also dynamical evidence that M31 (Andromeda)  has a compact
massive source at its center with $M\sim 3{\rm x}10^7 M_\odot$ which
favors the existence of a BH \cite{kormendy1}. 
M32 also harbors a central dark object
of $M\sim 5{\rm x} 10^6$. There is  evidence for very massive nuclei in
other nearby galaxies, in NGC 4594 (the Sombrero galaxy)
\cite{kormendy2} and in NGC 3115 \cite{kormendy3}.

The nuclei of most galaxies are inactive, in the sense that 
$L\sim 10^{-4}L_{\rm galaxy}$. There are active galactic nuclei (AGN) 
which can shine more than the entire galaxy. Galaxies that have AGN 
are $1\%$ of all the galaxies. 
Examples of AGN are the quasars (of which 3C273 has a luminosity equivalent 
to $10^3$ galaxies), blazars, Seyferts, radio galaxies and other objects. 
In a spectrocospic classification these AGN are divided in AGN type 1 
which show broad and narrow emission lines and AGN type 2 which show 
only narrow emission lines. 
They have some common features: (1) non-thermal radiation, 
(2) high concentration of mass in a small region, (3) variability 
in luminosity, (4) ejection of jets at great distances and 
(5) similarities with normal galaxies.  

The idea is to explain generically all different objects and phenomena
with one model. The most favored model invokes accretion onto central
supermassive BHs as the ultimate power source for these luminous objects
which radiate at the Eddingtom limit.  Even, if one invokes other central 
objects, such as spinars (yielding their rotational energy) or a cluster
of packed stars (supplying nuclear energy through supernova explosion),
the emission of so much energy from such a small volume (which is
measured through variability) leads inevitably to the collapse to a BH
\cite{rees1}. There is also the possibility that some galactic nuclei 
may contain two massive BHs in orbit around each other \cite{begel}. 
Further out from the central object, 
there is a dusty accretion torus which
provides a mechanism to understand AGN 1 and 2 \cite{antonucci}. The
jets, when they exist, point from the central region  into two opposite
directions aligned with the rotation axis of the torus. The blazars are
thought to be quasars with one jet pointing towards us. The
spectroscopic differences in AGN are also due to different orientations
of the torus with respect to the Earth.  If one can see the inner edge
of the torus one observes both the broad lines emitted in the inner region 
by high speed clouds and the narrow lines emitted in the outer 
edge. If the torus is seen edge-on only the narrow lines are observable.

The evidence to detect the central mass, 
both in AGN and normal galaxies, is based in most cases on the
increase of mass-to-light ratio 
in the central region. Only in two cases, M87 and NGC4258, is the value for
the central mass based on gas dynamics rotating around the central
mass. By using the Hubble Space Telescope it was possible to measure
the Doppler shift of emission lines from doubly ionized oxygen around
$R \sim 60$ly from the center of M87 \cite{ref3}.  This implies a
rotation velocity of $v\sim 550$Km/s for the gas in orbit which then
gives $M= \frac{v^2 R}{2G} \sim 2-3 {\rm x} 10^9M_\odot$. The mass is so
great in such a small region that is difficult to think of any other
explanation than a supermassive black hole inhabiting the center of the
galaxy. If, for instance, the mass were contained in solar type stars
in a dense cluster, they would be packed 100 thousand more times 
closely than
in the solar neighborhood. However, this is discarded, since there is
not enough light comimg from this region.  In the case of NGC4258,
recent work \cite{miyoshi} \cite{greenhill} has also pointed to the
confirmation of two things: 1. Keplerian velocities of $\sim 1000$Km/s in
an inner orbit of very small radius, $R\sim 0.4$ly, around the central
mass have been measured which imply a mass of $M \sim 2 {\rm x}
10^7M_\odot$. This work is considered to provide the strongest case for
a supermassive BH in the center confirming the predictions of
Lynden-Bell \cite{lynden-bell2}. 2. The velocities are measured through
water masers which are found to come from a torus-like region
confirming the unifying model of Antonucci and Miller
\cite{antonucci}.
It has been suggested \cite{kormendy4} that the best one can
do for the black hole case is to refute the other models on physical 
grounds.  For NGC4258 this has been undertaken \cite{maoz}.

One can thus have a model in which all galactic phenomena are unified,
not only within AGN themselves, but also relating normal galaxies and AGN.
In AGN, part of the potential energy is released
when matter approaches the event horizon and the energy escapes as
radiation providing the mechanism to power the emission. An accretion
rate of a few tens of $M_\odot$ per year, which can be supplied by
surrounding gas and by stars tidally disrupted in the gravitational
field, will provide a power greater than $10^{47}$ erg/s which would
explain even the highest quasar luminosities \cite{rees2}.
In normal galaxies there is no matter to be accreted.  The real
difference then between the nuclei of normal galaxies and AGN is, in
this model, not the mass of the BH, but the phase of the cycle of
accretion. The quiescent nuclei would be
BHs starved of fuel, i.e., dead quasars.

\vskip 3mm 

\noindent
{\bf 4. Quantum Black Holes and Elementary Particles}

\vskip 1mm

We have described in the previous sections stellar 
BHs with masses $3-20M_\odot$ (and $10-60$Km), and galactic BHs 
with $10^6-10^9M_\odot$. But there also exists the possibility 
of having BHs with much smaller masses. For instance, if the Earth 
with mass $\sim 10^{27}{\rm gm} \sim 10^{-6}M_\odot$ was 
compressed to a radius of $1$cm it would turn into a BH. There is 
the possibility that primordial BHs with mountain masses, 
$10^{14} {\rm gm} \sim 
10^{-19} M_\odot$, and a radius similar to the proton radius 
$10^{-13}$cm could be formed in the early universe \cite{carr1}. 
The smallest possible BH would have a mass of $10^{-5}$gm and 
a radius equal to $10^{-33}$cm, the Planck radius, which is thought to 
be the minimum possible radius that occurs in nature. Smaller 
masses would have a Schwarzschild radius smaller than the Planck 
radius, and thus if compressed into a BH, these masses would be 
snatched by the Planck regime, (i.e., by the spactime foam 
\cite{hawking1} \cite{wheeler}), before they had turned into a BH. 

Of course, these BHs have a totally different interest from the 
macroscopic and giant BHs. Their physical effects are of a different 
kind. Let us take the mountain mass BH, $M\sim 10^{-19} M_\odot$ and 
$R\sim 10^{-13}$cm. Its gravitational attraction at a distance of 
$10$m would be relatively small ($\sim 0.1$m/s$^2$). Its tidal force on a 
$1$cm tight object of $1$gm would barely be felt at a distance of 
$10$cm. 
Such a black hole could cause some damage on nearby objects, but not 
a lot. For a Planckian $10^{-5}$gm BH its gravitational atraction would 
give an acceleration of $10^{-6}$cm/s$^2$ at a distance of 
$10^{-5}$m, roughly the size of a living cell. 
On this basis, even if one of these BHs enters our body 
we would live without 
noticing it, the accretion onto it would be vanishingly small. 
However, there are other proceses that would made the BH noticeable, 
and these could cause damage in our body. 

There was a great turn in BH theory after Hawking found in 1974 that
BHs can radiate through quantum effects \cite{hawking2}.  There were
already hints that BHs have a thermodynamic behavior.  If one throws
entropy $S$ into the inside of the BH, this entropy disappears from our
universe in direct violation of the second law of thermodynamics. Since
there is a theorem \cite{hawking3}, within classical general
relativity, that states that in any process the area $A$ of a BH never
decreases, Bekenstein proposed that $S_{BH}\propto A$, such that the
second law is not violated, $S+S_{BH}\geq0$ always \cite{bekenstein}.  
Since, to an
entropy one can associate a temperature through the thermodynamic
relation $S=\frac{Q}{T}$, the BH must have a temperature. Indeed, by
complicated calculations of quantum field theory in a BH background,
Hawking was able to find  that the BH emits blackbody radiation at a
temperature $T= \frac{\hbar c^3}{8\pi G k_B}\frac1M \simeq 6 {\rm x}
10^{-8} (\frac{M_\odot}{M}) K$. Since so many fundamental constants
(section 1) enter this formula one can say that quantum mechanics,
general relativity and thermodynamics must merge together 
in a unified theory.
For $M\sim 1M_\odot$ one has $T\sim 10^{-7}$K. When $M\sim
10^{-19}M_\odot$ implies $T\sim 10^{12}$K. For a Planckian BH, 
$M\sim 10^{-5}$gm, therefore, $T\sim 10^{32}$K.  
Making the translation $E=k_B T$ one has that
the energies involved in the evaporation of a Planckian size BH are
$\sim 10^{19} {\rm Gev} \sim 100$watt-hour.  This could do some damage
in our body.  The mechanism for evaporation can be explained in several
ways, the most popular uses the idea that the vacuum is full of virtual
particles which are created and annihilated without violation of 
the uncertainty
relation $\Delta E\Delta t \buildrel>\over\sim \hbar$. However near the
event horizon it can happen that one particle enters the BH while the other
escapes out to infinity. The net result is blackbody radiation at the
Hawking temperature $T$. Now, the power emitted by a radiating BH
is $4\pi R^2 \sigma T^4=\frac{\lambda}{M^2}$, where $\sigma$ is the
Stefan-Boltzmann constant, and 
$\lambda= \frac{\hbar c^6}{15360\pi G^2}$, a value 
found through the equations given above.  Then, one finds that the BH
looses energy at a rate $\frac{dMc^2}{dt} = -\frac{\lambda}{M^2}$ which
can be integrated to give 
$M=({M_{\rm o}}^3 - 3\frac{\lambda}{c^2} t)^{\frac13}$. For 
an initial mass of $M_{\rm o} \sim 1M_\odot$ one gets that the BH
evaporates in $10^{67}$years.  If one puts
$M_{\rm o} \sim 10^{-19}M_\odot$ one finds $t\sim 10^{10}$ years, 
which means
if created in the primeval Universe these BHs should be evaporating by
now. This could happen in a burst of final radiation after passing
through the Planck scale. Some have speculated that the observed 
${\rm \gamma}$-ray
bursts could come from these mini-BHs, but there are tight limits on
their existence from gravitational lensing \cite{carr2}.

Thus, classically, BHs are stable, but 
quantum mechanically they are unstable, they slowly evaporate and shrink.
One striking effect that arises immediately is the violation of baryon 
number. Baryon number conservation is a law in elementary particle 
physics. However, if, say, a totally isolated neutron star of 
$10^{57}$ neutrons (baryons) collapses onto a BH, it will evaporate  
in a baryon-antibarion manner, actually most of the radiation will be 
in photons which carry zero baryon number anyway. Thus, gravity and 
quantum field theory produce violation of baryon number. 

One problem that Hawking radiation gives rise to is called
the information paradox. To describe a star completely, one must
specify a large ammount of information, such as, total mass $M$, total
charge $Q$, total angular momentum $J$, temperature, pressure,
gravitational multipole moments, other chemical potentials, and so on,
including the quantum states of the $10^{57}$ protons and neutrons that
constitute the star.  When the star collapses to form a BH, the no-hair
theorems say that the BH is described by only three parameters, $M$,
$Q$ and $J$. All the other information that was necessary to describe
the original star is now hidden inside the event horizon. Hawking
\cite{hawking4} found within his calculation, that the blackbody
thermal spectrum of the emitted flux of particles would not carry the
original information out to the exterior region. After the BH completely
evaporates,  the information that was trapped inside also vanishes with
the BH. The information paradox for BHs is the problem of explaining
what happens to the missing information.  It is of great importance
because, in usual quantum mechanics, the wavefunction $\psi$ evolves in
such a way that information contained in it is never lost. However, if
the picture described here is correct, then gravitational collapse
violates a fundamental principle of quantum mechanics.

Another important reason to study BH evaporation is that the final 
stages of the evaporation process involve physics near the Planck scale, 
where quantum gravity is expected to become important. Thus, BHs provide
a theoretical laboratory where one can gain insight into 
the physics at this minimum scale. 

All these issues are highly complicated in four spacetime dimensions.
To understand better these problems one must resort to lower dimension
theories. In two dimensions (one time plus one space dimension) general
relativity is trivial, it has no dynamics. However, if one adds a
dilaton scalar field the  theory has many features similar to four
dimensional general relativity (see, e.g.  \cite{lemossa1}).  There are
many  different theories in two dimensions with interesting dynamics
\cite{lemossa2}\cite{lemossa3}\cite{lemos2}.  One that has been
extensively studied \cite{witten} \cite{callan} \cite{horowitz} is
related to string theory (a consistent theory of quantum gravity, 
although it 
has problems in delivering the other three fundamental interactions).  
In three
dimensions general relativity has dynamics, although not much (the theory 
has no local degrees of freedom). Surprisingly, it has
been found that a three dimensional black hole in a space with constant
curvature exists \cite{banados1} \cite{banados2}.  One can connect
these three dimensional theories with four dimensional general
relativity \cite{lemos3} \cite{lemos4} 
\cite{sakleber} \cite{lemoszanchin}.  The
results obtained using two, three and four dimensional theories to solve
the information paradox are still controversial \cite{hooft}. 
However, theoretical
experiments, involving annihilation of a pair of BH-antiBH, have shown
that information can indeed disappear altogether, inside an event
horizon \cite{ross}.

Extreme BHs also provide interesting results. A charged BH is called 
extreme when $Q=M$ (in geometrical units where $G=c=1$, otherwise we 
can write $Q=\sqrt{G} M$). (If 
$Q>M$ then there is no horizon, instead one has a naked singularity, 
which if it exists complicates the thermodynamic picture. That is one 
reason why cosmic censorship \cite{penrose}, 
which forbidds the existence of naked 
singularities, is widely accepted). 
For extreme BHs the Hawking temperature is zero, they do not 
radiate. Thus they can be considered stable particles, that do not 
decay. If one of these BHs 
absorbs an infalling neutral particle, the BH's mass
will be increased, the charge-to-mass ratio is then lowered 
raising the Hawking temperature above zero. The BH then emits particles 
by the evaporation process, and returns to its ground state.  This 
appears as a scattering process, an incoming  initial state of one 
particle is scattered into 
other particles as a final state. 
There are other processes that resemble particle physics or
are connected to other physical branches, e.g.,   
BH-BH scatering, and the statistics a BH gas 
should obey (are BHs fermions, bosons or neither? \cite{strominger}).  

Physicists believe that gravity becomes the dominant 
field at the quantum Planck scale $10^{-33}$cm. It represents 
the minimum scale at which spacetime can be considered smooth. 
BHs are the objects to test this scale, through Hawking radiation, 
and related phenomena. Imagine the following futuristic experiment:
two incoming particles in a huge accelerator are set to collide 
face-on, such that, a center of mass energy of
$\sim 10^{19}$Gev is produced. Then, one might form a Planckian BH which will 
evaporate quickly in a burst, allowing us to study the physics at 
the Planck scale. One might think that by increasing the energy 
the study of sub-Planckian scales would follow. However, this is 
not the case. By increasing the energy one would produce a BH with 
larger mass, which would decay slowly, not allowing any test of 
Planckian physics.

\vskip 3mm

{\bf 4. Conclusions}
\vskip 1mm

BHs are used in many different phenomena, from high 
energy astrophysics to high energy elementary particle physics. 
The results brought from each area of study, either observational, 
theoretical or experimental, will serve to gain a better understanding 
of the physics of these beautiful objects. 
This review is a summary of some aspects of the nature of BHs. 

\vskip 6mm

\noindent 
{Acknowledgements} -- I thank Hor\'acio Dottori, Dalton Lopes and Vera 
Jatenco for inviting me to deliver the seminar to which this article 
corresponds, in the XXI$^{\rm th}$ SAB meeting, 1995. I thank Verne 
Smith for his careful reading of the manuscript.

\end{document}